\documentclass[sigconf, nonacm]{acmart}

\usepackage{algorithm}
\usepackage{algpseudocode}
\AtBeginDocument{%
  }

\begin{document}

\title{Dependencies and Dataflow in Seed-Filter-Extend Pipelines}

\author{Shiv Sundram}
\email{shiv1@stanford.edu}
\affiliation{%
  \institution{Stanford University}
  \city{Stanford}
  \state{California}
  \country{USA}
}

\renewcommand{\shortauthors}{Sundram}

\begin{abstract}
Comparing genomes is critical for discovering mutations, tracking evolutionary lineages, and advancing cross-species genomics. Fundamentally, this reduces to an $O(n^2)$ string-matching dynamic programming (DP) problem, a challenge that has driven decades of performance research. However, executing a strict $O(n^2)$ DP algorithm is computationally intractable for genomes spanning millions to billions of base pairs. Consequently, modern aligners rely on global heuristics to identify thousands of candidate similarity regions between species. Unfortunately, these methods are burdened by complex serial dependencies. Once candidate regions are identified, the pipeline executes localized DP alignments, which introduce their own non-trivial heuristics and irregular data dependencies. While parallelizing dense, two-dimensional DP is a well-studied problem, accelerating this end-to-end pipeline is significantly more challenging. Parallelizing across candidate regions and offloading irregular, heuristic-laden local alignments to modern hardware (such as GPUs) remains a major hurdle. In this work, we address the challenge of overcoming these serial bottlenecks by optimizing the global pipeline across regions. We take inspiration from four papers: LASTZ, SegAlign, Darwin-WGA, and SNAP, synthesizing findings across each to inform optimizations, which we either prototype or implement directly in LASTZ.
\end{abstract}

\received{2 June 2026}

\maketitle

\section{Seed-Filter-Extend Algorithms}
The computational challenge of genomic alignment is fundamentally rooted in string-matching dynamic programming (DP), a historically $O(n^2)$ problem. Accordingly, there is no shortage of libraries implementing Smith-Waterman and Needleman-Wunsch dynamic programs for aligning sequences \cite{sundram2024compiling, tariq2025reptile, daily2016parasail, vinaithirthan2025code}. However, to render this computation tractable for sequences spanning millions to billions of base pairs, the bioinformatics community shifted toward heuristics, a paradigm pioneered by tools like BLAST \cite{madden2013blast}. Rather than computing a full DP matrix, these algorithms identify exact or near-exact short sequence matches (seeds) to serve as anchors, localizing the computationally expensive DP extensions only to regions of high probable homology.

Accordingly, a common computational motif has emerged among modern tools for sequence alignment: the widespread adoption of the seed-filter-extend pipeline. This multi-stage approach first discovers small, exact or near-exact matches (seeds) to serve as initial candidate regions of potential homology. A localized dynamic programming (DP) algorithm is then executed strictly at these candidate regions, and the resulting alignments are subsequently filtered and chained. This architecture is not merely a performance optimization to bypass an intractable $O(n^2)$ global search; it is biologically necessary for cross-species genomics. Over deep evolutionary time, genomes undergo massive structural variations, meaning subregions may be translocated, inverted, or duplicated entirely. While traditional local DP algorithms (like Smith-Waterman) can theoretically identify isolated regions of similarity, they lack the macro-level logic to associate them. The seed-filter-extend paradigm—particularly through its chaining phase—elegantly handles these repetitions and scramblings, providing the structural flexibility required to map complex evolutionary events across divergent species.

\section{State of the Art Genomic Alignment Tools }

Within this seed-filter-extend paradigm, different tools have evolved to address highly distinct biological problem spaces. LASTZ \cite{harris2007improved} represents the gold standard for deep evolutionary, whole-genome comparisons (e.g., aligning human and mouse genomes). Because evolutionary divergence introduces significant substitutions, insertions, and deletions, LASTZ prioritizes maximum sensitivity. It achieves this by employing short, spaced seeds designed to tolerate a high frequency of transition mutations. While LASTZ's heuristics are biologically rigorous, they are deeply rooted in a CPU-centric execution model. Recent efforts, such as SegAlign \cite{goenka2020segalign}, have attempted to accelerate LASTZ by porting its workload to modern GPU architectures; however, SegAlign fundamentally inherits the original LASTZ algorithmic pipeline, carrying forward its structural limitations. 

SegAlign specifically targets the intermediary filter stage—the ungapped extension (X-drop)—for parallelization. Because the vast majority of initial seed hits in complex genomes are false positives, they are discarded during this phase. SegAlign offloads this entire filtration workload to the GPU, an architectural decision grounded in three key factors: first, each seed hit can be filtered independently; second, the evaluation of a single hit maps elegantly to a GPU warp, which can compute the filtration systolically; and third, in deep cross-species comparisons (where genomes vary significantly), this filtering stage overwhelmingly dominates the total execution time. However, as will be demonstrated later in this report, this fundamental assumption breaks down when comparing highly similar genomes (e.g., within primate lineages). In high-identity regimes, the computational bottleneck shifts dramatically, and the gapped extension (Y-drop) phase dominates the runtime. Therefore, a primary contribution of this work is the acceleration of this heavily bottlenecked extension phase, overcoming its serial dependencies and using a systolic technique (adapted from Darwin-WGA, a LASTZ ASIC \cite{turakhia2019darwin}) for computing the extensions on a GPU warp.

In stark contrast to whole-genome aligners, tools like SNAP \cite{zaharia2011faster} were engineered to manage the explosion of high-throughput, short-read sequencing data. SNAP operates under a fundamentally different biological assumption: the sequences being aligned represent high-identity reads from the same (or a highly similar) species. Consequently, SNAP trades deep sensitivity for extreme specificity and throughput. It utilizes much longer, contiguous seeds (often 20 or more bases) and vast memory-resident hash tables. This architectural divergence highlights a critical dichotomy in modern bioinformatics: legacy aligners possess the sensitivity required for complex whole-genome assembly, while modern read mappers possess the hardware sympathy required for high-throughput acceleration.

\section{Bottleneck Analysis of Seed-Filter-Extend in LASTZ}

\begin{figure}
    \centering
    \includegraphics[width=0.99\linewidth]{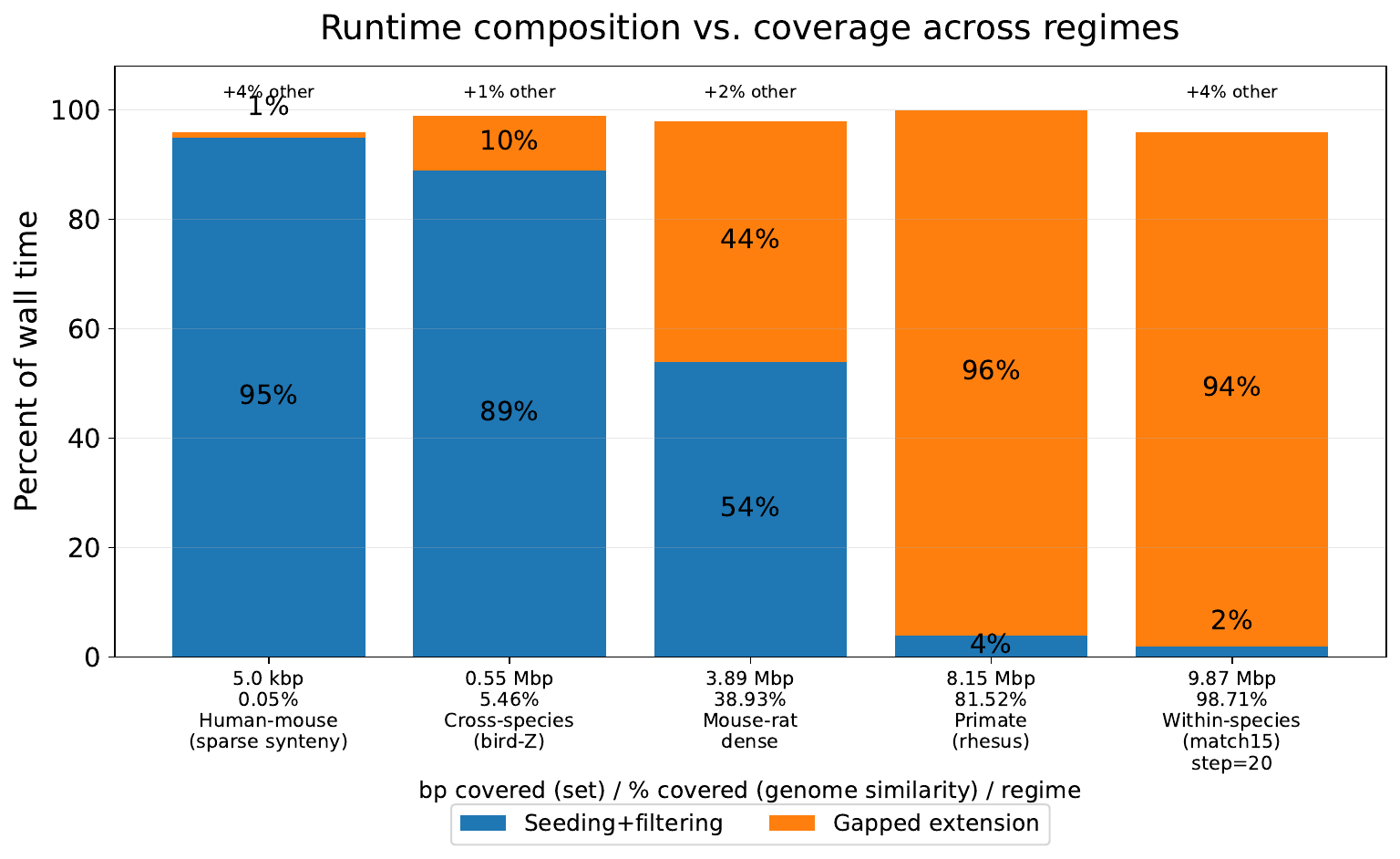}
    \caption{Distribution of runtime across the seed/filter stage and extension stage in LASTZ. Filtering, the traditional target of parallelization, dominates when species have vastly different genomes. However, once genomes approach a level of similarity, runtime is almost entirely monopolized by the relatively serial extension stage.}
    \label{fig:rundistr}
\end{figure}

While LASTZ remains the gold standard for cross-species sequence alignment, prior efforts to parallelize its pipeline (such as SegAlign) have predominantly targeted the initial ungapped extension (X-drop) filtering stage due to its naturally uncoupled dataflow; each seed hit can be filtered (via X-drop) in parallel. Figure~\ref{fig:rundistr} illustrates the runtime distribution between the X-drop filtering and gapped extension (Y-drop) phases across varying degrees of evolutionary divergence. In highly divergent comparisons (e.g., human versus mouse), the vast majority of execution time is consumed by X-drop filtering, effectively validating SegAlign's architectural focus. However, as the compared genomes converge in similarity (e.g., intra-primate alignments), the computational bottleneck shifts dramatically; nearly the entire runtime is monopolized by the Y-drop extension phase. The rat-mouse comparison highlights a critical inflection point where moderate synteny results in an equal distribution of compute across both phases. Consequently, while GPU-accelerated filtering provides a vital foundation for divergent taxa, whole-genome alignments within high-identity regimes remain severely throttled by legacy serial dependencies. To resolve this structural inefficiency, our optimization efforts directly target the parallelization of the Y-drop extension stage, which is computationally expensive but relatively serial.

\section{Architectural Limitations and Algorithmic Deep Dive}

The difficulty of accelerating whole-genome alignment on modern SIMD (Single Instruction, Multiple Data) hardware lies in the structural dependencies of the underlying algorithms. LASTZ's pursuit of sensitivity incurs a massive computational penalty. Its lightweight, spaced seeds trigger a high volume of false-positive candidate regions in complex genomes. The aligner spends the vast majority of its execution budget evaluating these candidates via ungapped extensions (X-drops), only for most to fail.

More critically, when LASTZ does initiate the intensive gapped extension phase (Y-drop), it enforces strict serial heuristics to prevent redundant calculations (Algorithm \ref{alg:lastz}). As the algorithm traverses the genome, it maintains dynamic data structures---specifically, segment lists to track the boundaries of aligned regions. Before a new seed can be extended, LASTZ performs containment skipping and neighbor-masking to verify that the new anchor does not overlap with a previously resolved alignment. This results in an almost 10x reduction of work compared to LASTZ with containment skipping turned off \cite{sundram2024lastz}. However, it also creates a severe loop-carried dependency: the evaluation of alignment $N$ cannot proceed until the spatial boundaries of alignment $N-1$ are definitively computed. This serial trap renders the gapped extension phase highly resistant to parallelization. Accelerator frameworks like SegAlign struggle against this bottleneck, as modern GPUs require massive, uninterrupted batches of independent work to keep their multiprocessors saturated.

\begin{algorithm}[ht]
\caption{Legacy LASTZ (Strict Serial)}
\label{alg:lastz}
\begin{algorithmic}[1]
\For{$j \gets 0$ \textbf{to} $N-1$}
    \State $s \gets \text{Seeds}[j]$
    \If{\text{IsMasked}(s) \textbf{or} \text{IsContained}(s)}
        \State \textbf{continue}
    \EndIf
    \If{\text{IsFiltered}(s)}
        \State \textbf{continue}
    \EndIf
    \State $a \gets \text{YDropExtension}(s)$ \Comment{Heavy serial compute}
    \If{$\text{Score}(a) \geq \tau$}
        \State \text{CommitAlignment}(a)
        \State \text{UpdateMaskingLists}(a) \Comment{Loop-carried dependency}
    \EndIf
\EndFor
\end{algorithmic}
\end{algorithm}

SNAP bypasses this serial trap entirely because of the nature of its input data (Algorithm \ref{alg:snap}). In short-read mapping, every sequencing read is mathematically and spatially independent. The alignment of one read has no bearing on the alignment of the next. SNAP exploits this uncoupled dataflow by loading massive batches of reads and evaluating their edit distances concurrently against the reference hash table. There is no shared state to update and no sequential masking required. The architectural convergence problem is therefore clear: to achieve SNAP-like throughput for whole-genome comparisons, an aligner must shed the strict serial dependencies of LASTZ without sacrificing the biological integrity of its outputs.

\begin{algorithm}[ht]
\caption{SNAP (Independent High-Throughput)}
\label{alg:snap}
\begin{algorithmic}[1]
\ForAll{read $r$ \textbf{in} Batch} \Comment{Parallel execution}
    \State Candidates $\gets \text{HashTableLookup}(r)$ \Comment{Highly specific seeds}
    \State $\text{best\_a} \gets \emptyset$
    \ForAll{$s$ \textbf{in} Candidates}
        \State $a \gets \text{EditDistanceExtension}(s)$ \Comment{Independent compute}
        \State $\text{best\_a} \gets \max(\text{best\_a}, a)$
    \EndFor
    \State \text{WriteOutput}(\text{best\_a}) \Comment{No shared state/masking}
\EndFor
\end{algorithmic}
\end{algorithm}

\section{Speculative Execution: A Hybrid Architecture for Batched Parallelism}

To break the performance ceiling of legacy whole-genome alignment, this work introduces a hybrid architecture that forces SNAP-like batched parallelism into the inherently serial LASTZ pipeline (Algorithm \ref{alg:speculative}). This approach specifically targets the ``high-identity'' regime of genomic comparison. By artificially shifting LASTZ's parameters toward longer, more specific seeds in these dense regions, we drastically reduce the false-positive extension penalty, exposing a smaller, higher-quality set of candidate anchors suitable for batch processing.

The core algorithmic contribution is the implementation of \textit{speculative execution} to artificially manufacture data independence. Rather than stalling the pipeline to update and query masking lists sequentially, our architecture temporarily suspends these serial heuristics. This isolates the inner DP cell loop, allowing the workload to be restructured into a three-phase batched pipeline:

\begin{enumerate}
    \item \textbf{Serial Scan:} The CPU rapidly iterates through the position table, gathering a fixed batch of high-quality seeds that are not masked by previously committed, historical alignments.
    \item \textbf{Parallel Y-Drop:} This batch is offloaded to the GPU, where thousands of independent, speculative Y-drop extensions are executed concurrently across wavefronts.
    \item \textbf{Serial Commit:} The GPU results are returned to the CPU, where overlapping boundaries and intra-batch containments are resolved in a lightweight post-processing step. Redundant alignments are filtered, and the surviving, high-scoring alignments are committed to the formal masking lists.
\end{enumerate}

\begin{algorithm}[ht]
\caption{Ours: Speculative LASTZ (Batched Dataflow)}
\label{alg:speculative}
\begin{algorithmic}[1]
\While{$j < N$}
    \State Batch $\gets \emptyset$
    \While{$|\text{Batch}| < K$ \textbf{and} $j < N$} \Comment{Phase 1: Serial scan}
        \If{\textbf{not} \text{IsMasked}(\text{Seeds}[j])}
            \State Batch $\gets \text{Batch} \cup \{\text{Seeds}[j]\}$
        \EndIf
        \State $j \gets j + 1$
    \EndWhile
    
    \ForAll{$s \textbf{ in} \text{ Batch }$} \Comment{Phase 2: Parallel Y-Drop}
        \State $\text{Ext}[s] \gets \text{YDropExtension}(s)$ \Comment{Speculative compute}
    \EndFor
    
    \For{ $s \textbf{ in } \text{ Batch}$ } \Comment{Phase 3: Serial commit}
        \If{\textbf{not} \text{checkContain}(\text{Ext}[s]) \textbf{and} $\text{Score}(\text{Ext}[s]) \geq \tau$}
            \State \text{CommitAlignment}(\text{Ext}[s])
            \State \text{UpdateMaskingLists}(\text{Ext}[s]) \Comment{Deferred state update}
        \EndIf
    \EndFor
\EndWhile
\end{algorithmic}
\end{algorithm}

\begin{figure}
    \centering
    \includegraphics[width=0.99\linewidth]{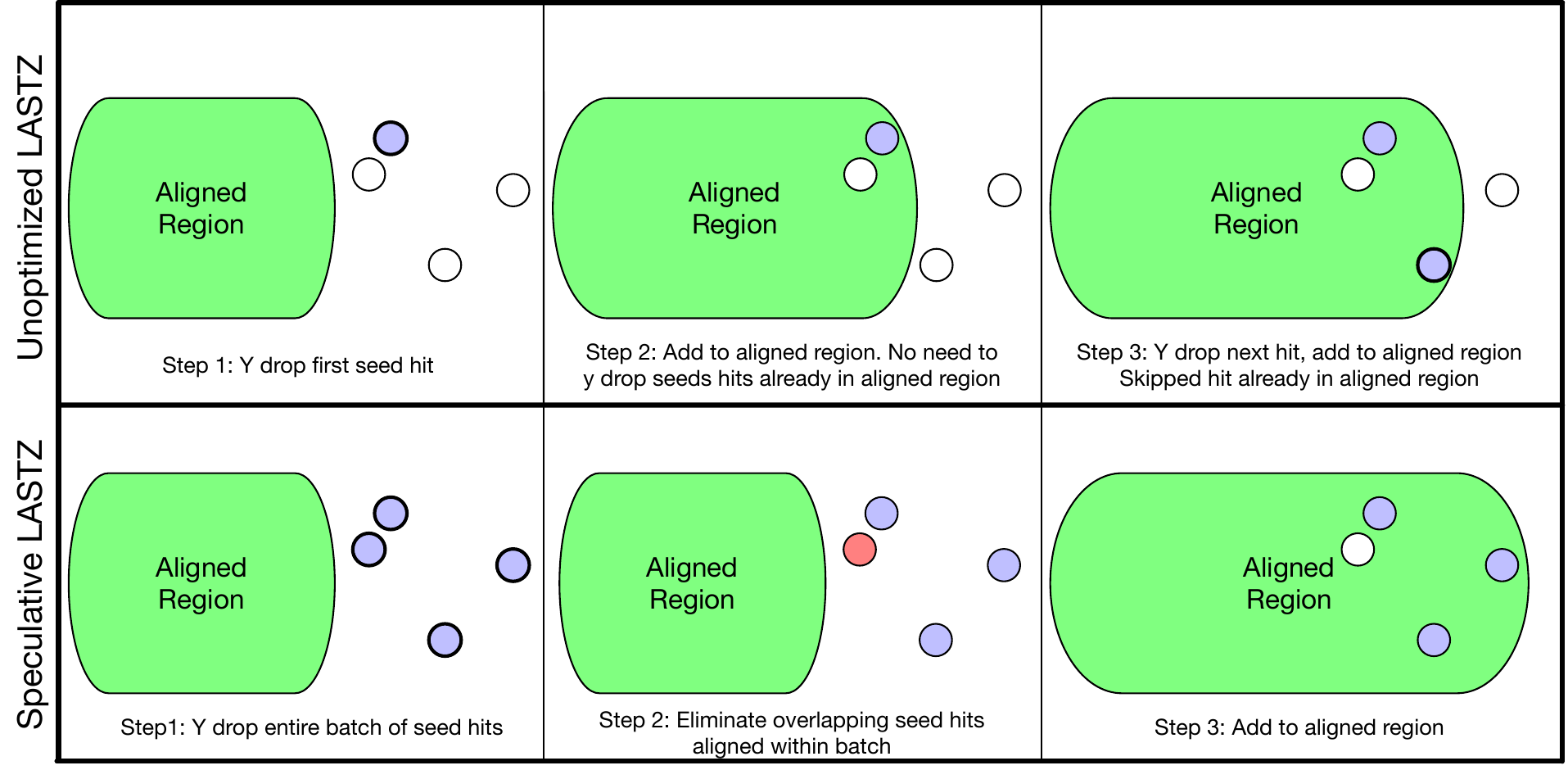}
    \caption{Speculative vs non-speculative execution of LASTZ}
    \label{fig:pdeferred}
\end{figure}

Ultimately, this architecture represents a deliberate, hardware-aware trade-off. Figure~\ref{fig:pdeferred} illustrates how the speculative execution differs from LASTZ's baseline execution model. By deferring containment and masking checks, the system will inevitably compute redundant gapped extensions on the GPU that are subsequently discarded during the serial commit phase. However, in the context of massive SIMD throughput, sacrificing raw compute cycles is vastly cheaper than stalling the pipeline to resolve serial branches. By engineering this speculative independence, our hybrid algorithm successfully maps an $O(n^2)$ dependency workload onto modern parallel architectures, achieving massive throughput acceleration while strictly preserving the biological accuracy of the legacy LASTZ output.

\section{Evaluation of Speculative LASTZ}

\begin{figure}
    \centering
    \includegraphics[width=0.99\linewidth]{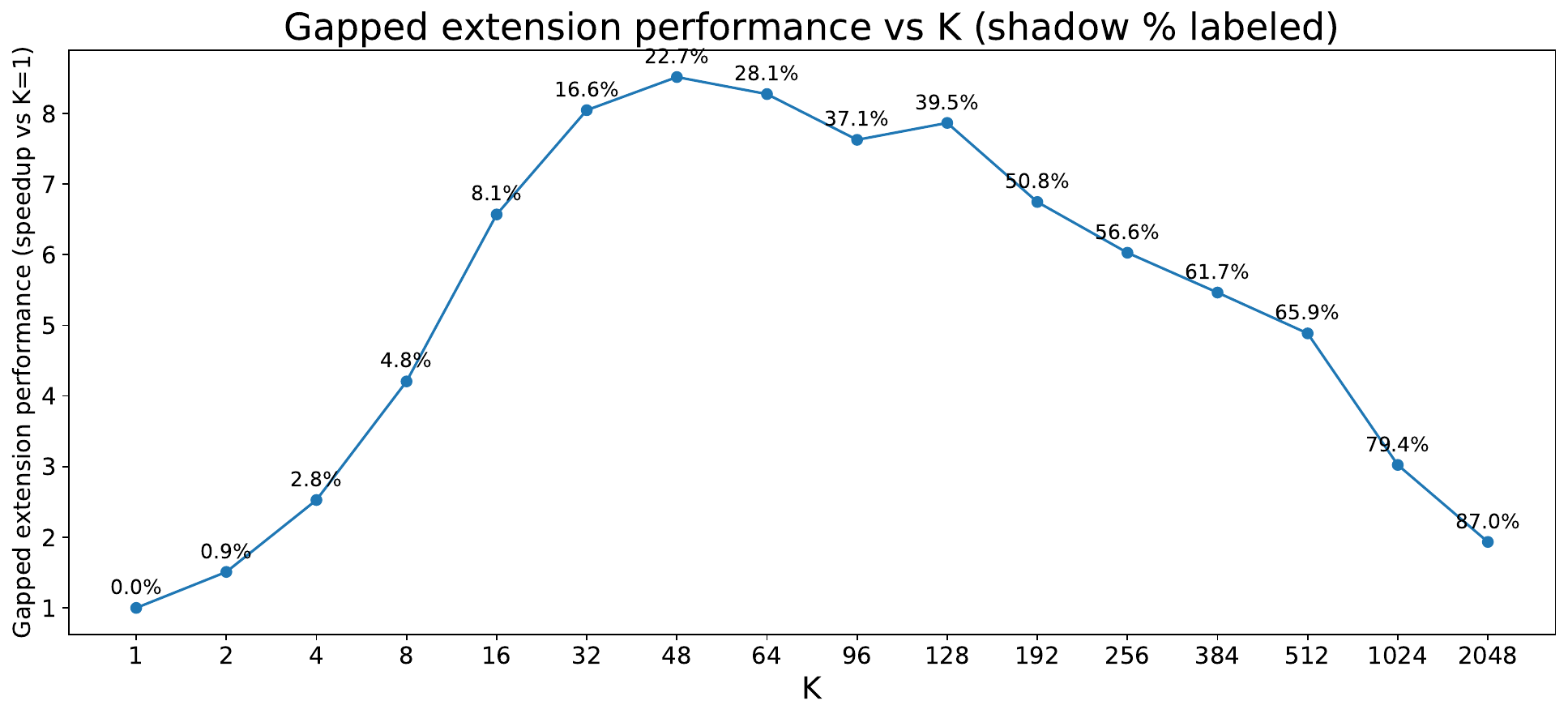}
    \caption{Increasing batch size (K) for the speculative step increases the potential parallelism. It also increases the percentage of work that becomes wasted shadow work. A batch size between K=32 and 64 achieves a sweet spot.}
    \label{fig:shadow}
\end{figure}

By definition, speculative execution introduces computational overhead, preventing perfectly linear scaling as thread counts increase. To optimize this architecture, two primary hyperparameters must be tuned: the thread count and the speculative batch size ($K$). While expanding the batch size exposes greater parallelism to the hardware, it simultaneously increases the frequency of overlapping alignments. This redundant computation, termed ``shadow work,'' is ultimately discarded during the serial commit phase. Figure~\ref{fig:shadow} illustrates the impact of varying batch sizes on overall performance for a fixed core count. When plotted on a logarithmic scale, the performance profile exhibits a distinct bell-shaped curve, revealing a critical architectural trade-off. We empirically identified an optimal batch size of $K=48$, which maximizes hardware saturation while keeping the penalty of shadow work within acceptable bounds.

To evaluate multi-core scaling and facilitate a simpler parallel prototype, we engineered a functionally equivalent alternative to LASTZ's gapped extension algorithm. We implemented all code in our own fork of LASTZ \cite{sundram2024lastz}. Standard LASTZ is highly memory-optimized, utilizing linear-space arrays (e.g., three $O(n+m)$ buffers) to evaluate a sequence subregion. To simplify parallel offloading, our prototype allocates the full $O(n \times m)$ dynamic programming submatrix. Due to this increased memory and cache footprint, our unoptimized, single-threaded prototype executes approximately $3\times$ slower than the highly tuned LASTZ baseline. However, the architectural advantages of speculative batching quickly eclipse this single-thread deficit. When scaled to 32 cores, our parallel implementation achieves an $8\times$ speedup over its own single-threaded baseline, ultimately outperforming the legacy LASTZ execution by $2.5\times$. This suggests a massive performance ceiling: if LASTZ's memory-efficient linear-space optimizations were retrofitted into our speculative parallel framework, the overall speedup would likely multiply, further separating this hybrid architecture from legacy limitations.

\begin{figure}
    \centering
    \includegraphics[width=0.99\linewidth]{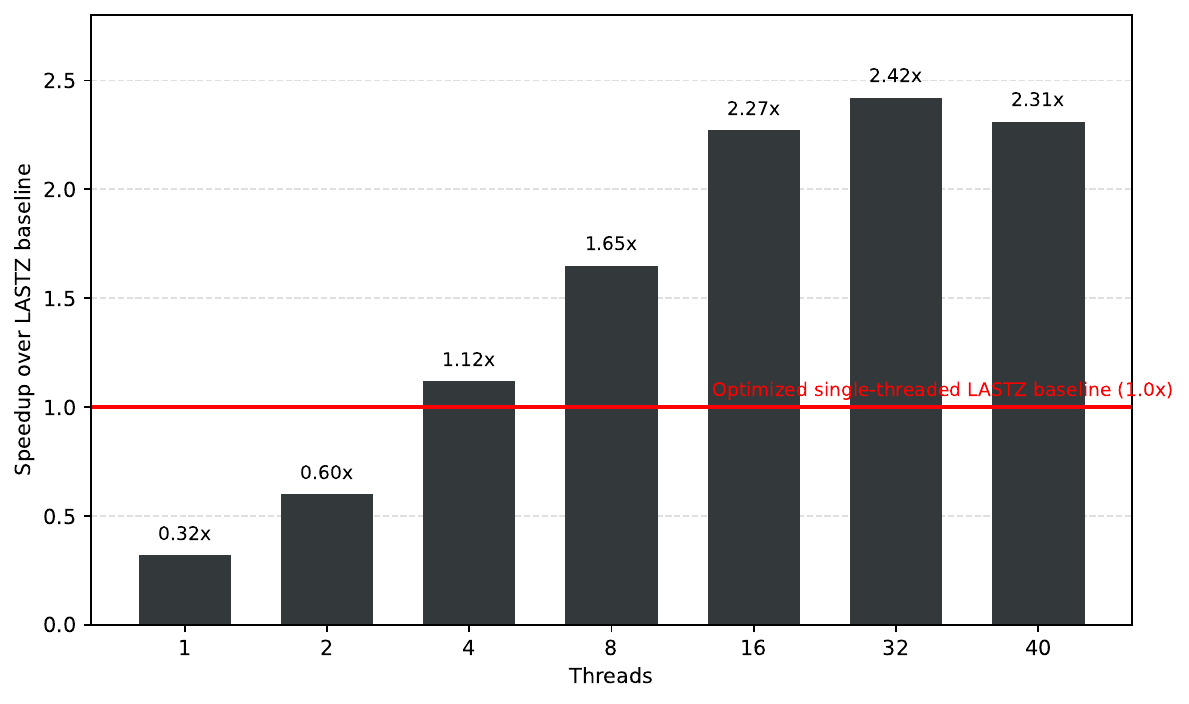}
    \caption{Speedup of speculative execution over LASTZ baseline}
    \label{fig:speedup}
\end{figure}

\section{Adapting High-Identity Seeding Heuristics}

Beyond parallelizing the extension dataflow, we adapt a secondary heuristic from SNAP to further optimize the LASTZ pipeline. SNAP was engineered specifically for short-read mapping, a domain characterized by extremely high sequence identity (e.g., mapping human reads to a human reference genome, which share $>99\%$ similarity). Because of this high identity, SNAP can utilize much longer, heavier seeds (e.g., 20 or more contiguous bases). In traditional, deep-evolutionary cross-species comparisons, heavy seeds severely degrade aligner sensitivity; they require long stretches of exact homology and exhibit little tolerance for transition or transversion mutations. However, the SNAP authors demonstrated that in high-identity environments, heavy seeds are biologically permissible. By enforcing strict specificity at the seeding stage, the algorithm drastically reduces the volume of initial seed hits, thereby eliminating the computational cost of filtering and extending millions of doomed false-positive alignments.

\begin{figure*}[t]
    \centering
    \includegraphics[width=0.99\textwidth]{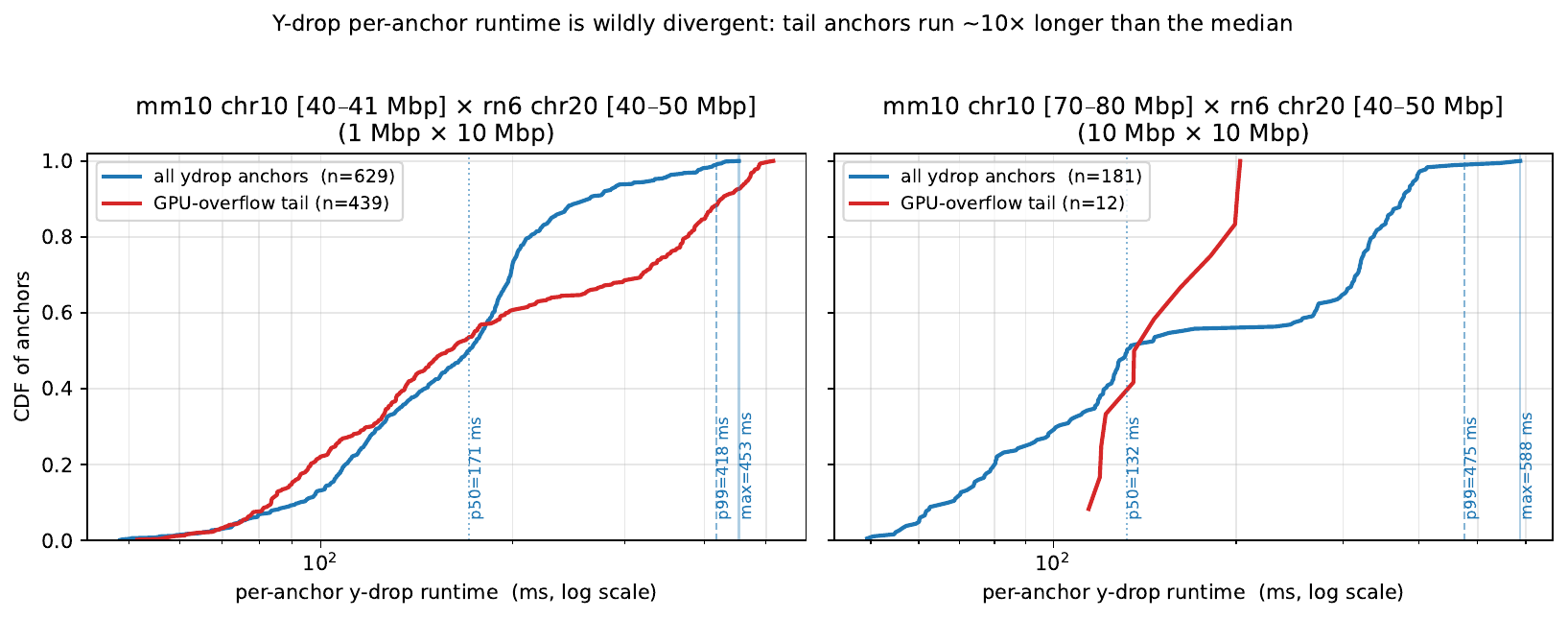}
    \caption{Divergence of extension times for seed hits. Extensions that overflow GPU memory  spill to the CPU (red line)}
    \label{fig:divergent}
\end{figure*}

\begin{figure}
    \centering
    \includegraphics[width=0.99\linewidth]{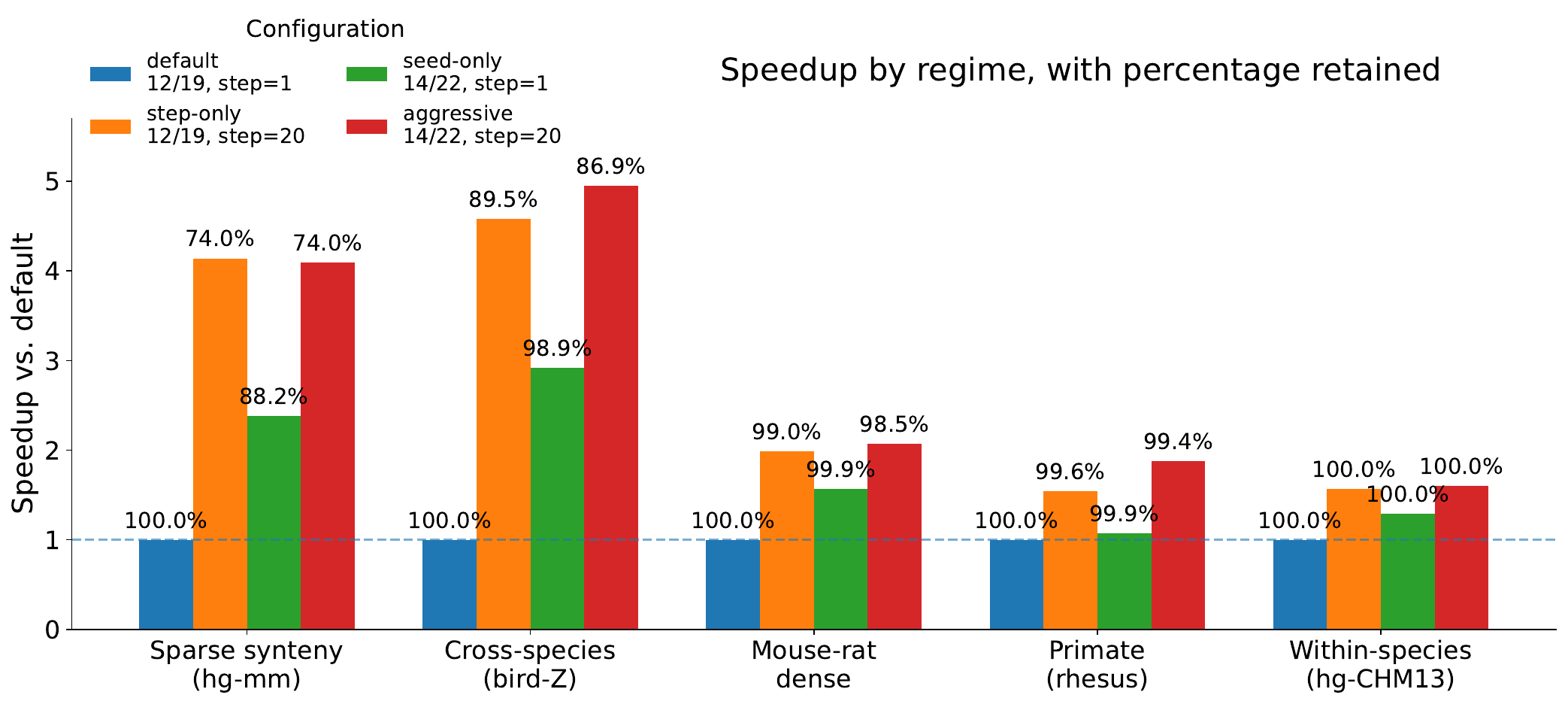}
    \caption{Result of increasing seed size and seed spacing ("step") on alignment speedup, shown as speedup bars, and on alignment sensitivity vs LASTZ's size=12 step=1 seed baseline.}
    \label{fig:seeds}
\end{figure}

We hypothesized that this high-identity seeding strategy could be successfully ported into LASTZ's whole-genome cross-species pipeline, provided the compared species exhibit sufficient evolutionary proximity (e.g., intra-primate comparisons). 

By default, LASTZ employs an advanced spaced seed pattern of weight 12 and span 19 (often denoted "12 out of 19"). This means it applies a predefined 19-bit mask where 12 specific positions are set to 1, requiring exact character matches between the target and reference only at those 12 positions. This empirically derived mask (1110100110010101111) affords the flexibility and sensitivity to transition mutations necessary for deep cross-species comparisons. In contrast, SNAP achieves massive throughput gains by utilizing longer, contiguous exact matches (e.g., size 20) to reduce the downstream computational load. We hypothesized that by adopting a heavier, SNAP-inspired spaced seed (specifically, a "14 out of 22" mask) in high-identity comparisons, we could produce high-quality alignments without a significant loss in accuracy.

Our empirical evaluations confirm this hypothesis. As illustrated in Figure~\ref{fig:seeds}, increasing the stringency of LASTZ's seeding parameters---specifically, increasing the seed weight from 12 to 14 and expanding the inter-seed spacing (the "step size")---drastically reduces the candidate pool and accelerates the entire pipeline. Crucially, this optimization has a negligible impact on aligner accuracy; the total number of correctly aligned base pairs between the target and query sequences remains practically unchanged for similar species. The heavier seed achieves 99\% of the accuracy of the baseline seed while doubling overall performance. By artificially suppressing the false-positive generation rate at the top of the pipeline, this SNAP-inspired heuristic acts as a powerful performance multiplier for our accelerated gapped-extension architecture.

\section{Challenges in GPU Y-Drop Acceleration and Future Work}

To contextualize the difficulty of parallelizing the gapped extension (Y-drop) phase, it is instructive to examine the acceleration strategies of modern tools like Darwin-WGA and SegAlign. Both architectures successfully target the ungapped filtering step (X-drop) by exploiting its massive, inherent parallelism. Because X-drop relies on a 1-dimensional dynamic program without gap penalties, its computational footprint is highly predictable. Each candidate seed is mapped to a GPU warp, which computes the extension systolically across threads. This results in an optimal, load-balanced execution model: there is ample independent work to saturate the GPU multiprocessors, and the computational load remains uniformly distributed across warps. 

We initially hypothesized that Darwin-WGA's systolic DP strategy could be adapted to accelerate the 2-dimensional Y-drop phase across GPU warps. However, our empirical evaluations revealed that this direct port yields no tangible runtime improvement due to severe load imbalance. Unlike X-drop, the Y-drop algorithm possesses a highly dynamic and unpredictable termination condition (when the alignment score drops below a specific threshold). Consequently, execution lengths are wildly divergent across different seed hits. When mapped to a synchronous GPU warp, this divergence triggers catastrophic load imbalance; the execution time of the entire warp becomes strictly bottlenecked by the single longest extension within that batch, leaving the remaining threads idle. Because the total size of a Y-drop extension cannot be known a priori, static load balancing is impossible.

This warp divergence penalty is further exacerbated by the architectural constraints of our speculative pipeline. While speculative execution successfully extracts artificial parallelism from the legacy LASTZ pipeline, the absolute volume of concurrent work remains strictly limited by the batch size to avoid excessive shadow work. We validate this empirically. On the mm10 chr10 × rn6 chr20 (mouse vs. rat) workload, per-anchor Y-drop runtimes span 2.5–4.5× from median to maximum (Figure~\ref{fig:divergent}). Because every block in our kernel maps to one anchor-half and waits on the slowest extension in its warp, this multiplier is the floor on within-batch GPU efficiency: even with perfect work-stealing the longest extension serializes the launch. 

\begin{figure*}[t]
    \centering
    \includegraphics[width=0.99\textwidth]{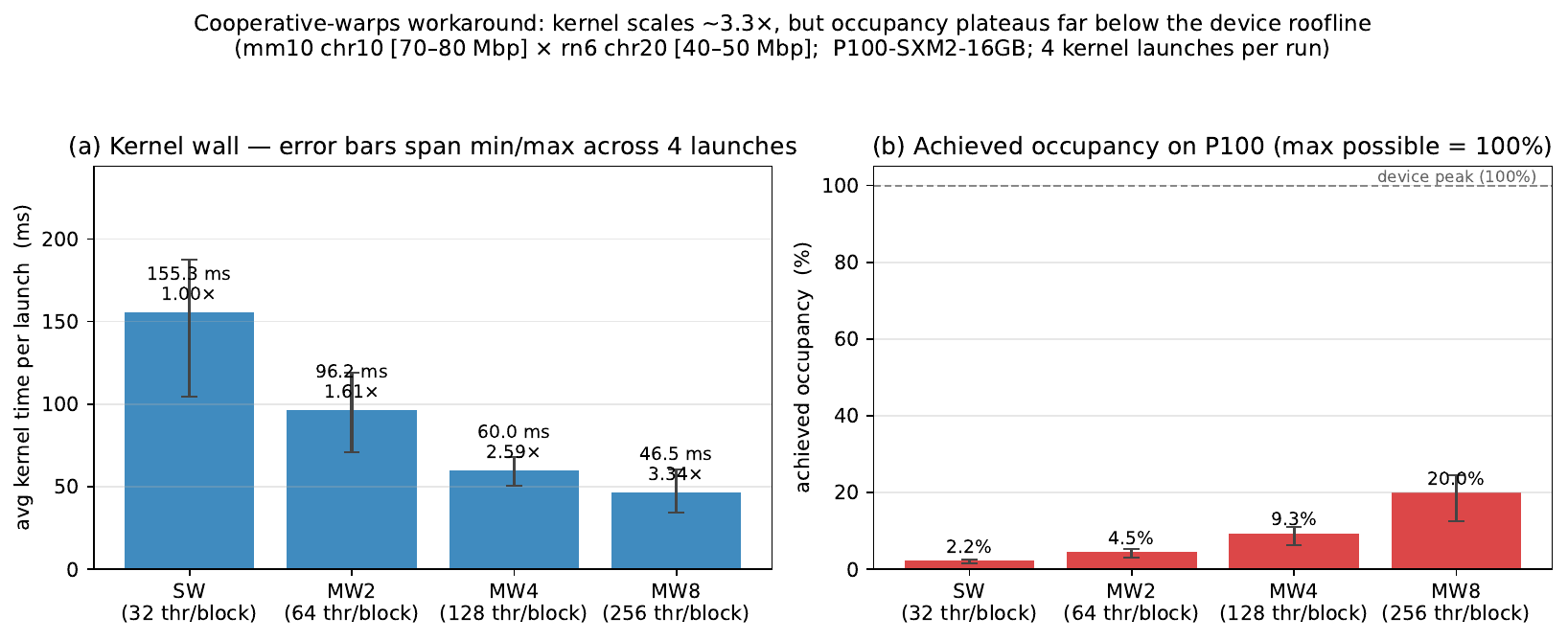}
    \caption{Runtime of systolic GPU y-drop kernel}
    \label{fig:scan}
\end{figure*}

\begin{figure}[t]
    \centering
    \includegraphics[width=.6\linewidth]{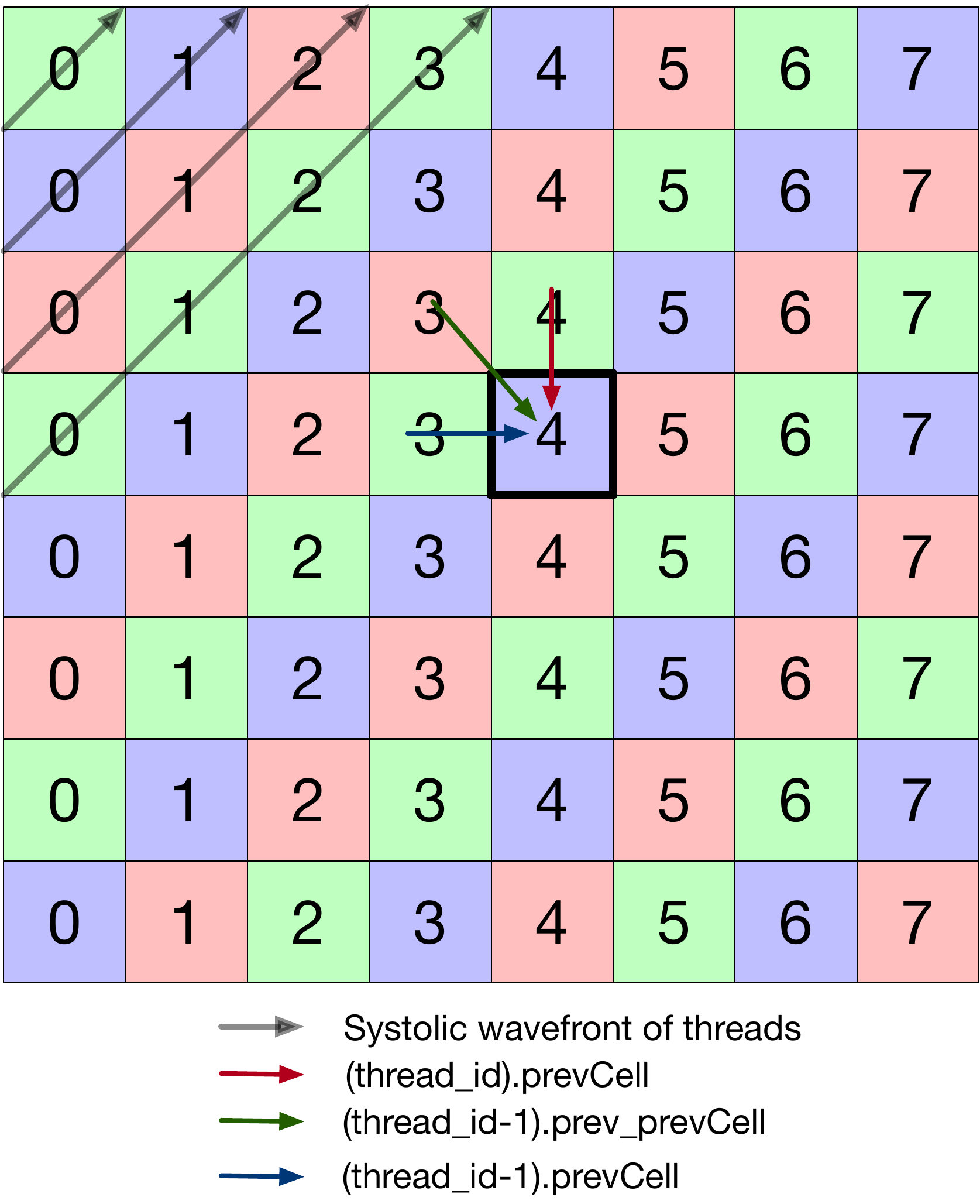}
    \caption{GPU thread layout (systolic)}
    \label{fig:systolic2}
\end{figure}
Furthermore, at our empirically determined optimal batch size (K=64), our systolic-warp kernel achieves only 2.2\% of the P100's theoretical occupancy (Figure~\ref{fig:scan}). A cooperative-warps remapping that uses 8 warps per anchor-half lifts achieved occupancy to 20\% and shrinks kernel wall time by 3.3×, but the ceiling remains five-fold below device peak: K=64 batches of speculative work simply do not generate enough independent extensions to saturate 56 SMs with 64-warp capacity. Ultimately, the speculative batch size does not provide enough concurrent work to saturate the device, meaning the hardware cannot rely on massive thread concurrency to hide the latency of divergent warps.

Developing a strategy to handle this extreme performance dynamism remains a critical area for future work. Modern high-throughput tools like SNAP mitigate workload disparity by pre-sorting sequence reads by length and deploying persistent thread queues, which dynamically distribute equally-sized slices of work. However, SNAP's strategy relies on two conditions absent in whole-genome alignment: fixed, predetermined query sizes, and a massive pool of millions of independent tasks. Engineering a dynamic, hardware-aware load-balancing strategy capable of managing highly divergent, unpredictable DP workloads within small, speculative batches ($K \le 64$) is the necessary next step to fully realizing the potential of GPU-accelerated whole-genome alignment.

Our systolic GPU kernel itself, however, can still be useful. The authors of Darwin-WGA note that using the 2D y-drop for the filtering stage, instead of using the 1D version (x-drop) employed by SegAlign for filtering, results in an increase in accuracy. Figure~\ref{fig:systolic2} illustrates the design of systolic GPU alignment kernel inspired by Darwin-WGA's custom systolic array. Specifically, the figure illustrates the thread to cell mapping, along with the thread wavefronts, for a 2D alignment matrix. At each stage, a particular thread only needs to read its own value from a previous wavefront, as well as the previous thread's values from the past 2 wavefronts. These values from the previous thread can be read by the current thread using GPU warp shuffle intrinsics, thus creating an inter-thread systolic communication pattern that eliminates the need for shared memory (L1 cache), which is the standard but slower method of communicating values amongst threads in a warp. This strategy results in a kernel that has minimal L1 cache requirements and is >2x faster than a kernel reliant on global memory to perform inter-thread communication. Replacing the actual LASTZ x-drop filter kernel with y-drop and measuring the differences in performance and sensitivity is left as important future work.

\section{Conclusion}

In this work, we addressed a critical computational bottleneck in modern sequence alignment: the size of the seed, and particularly the strict serial dependencies of the gapped extension (Y-drop) phase within the seed-filter-extend paradigm. While previous acceleration efforts have successfully parallelized the ungapped filtering stages for divergent genomes, we demonstrated that whole-genome alignments in high-identity regimes are severely throttled by legacy loop-carried dependencies. 

To break this performance ceiling, we introduced a hybrid architecture that incorporates speculative execution into LASTZ. By artificially deferring sequential containment and masking checks, we manufactured data independence, allowing an inherently $O(n^2)$ dependency workload to be restructured into a batched, parallel pipeline. Our evaluations showed that with an optimal speculative batch size, this approach overcomes the penalty of redundant ``shadow work,'' achieving up to a 2.5$\times$ speedup over the highly-tuned LASTZ baseline while strictly preserving its biological sensitivity. We further amplified these gains by adapting high-identity seeding heuristics from short-read mappers like SNAP, proving that heavier, widely-spaced seeds can drastically suppress false-positive extensions in cross-species whole-genome comparisons without sacrificing accuracy.

Finally, we highlighted the persistent challenges of migrating irregular dynamic programming workloads to modern SIMD hardware. Our attempts to port the Y-drop phase to a systolic GPU architecture revealed that unpredictable execution lengths and small speculative batch sizes lead to catastrophic warp divergence and poor device occupancy. Overcoming this extreme performance dynamism through hardware-aware, dynamic load-balancing remains a critical frontier. Furthermore, our highly optimized, cache-efficient systolic 2D Y-drop kernel opens promising avenues for future research, particularly in evaluating its potential to replace traditional 1D X-drop filtering to yield higher alignment accuracy. Ultimately, bridging the gap between legacy biological sensitivity and modern high-throughput hardware requires rethinking the fundamental dataflow of genomic aligners, and our speculative architecture provides a foundational step in that direction.
\bibliographystyle{ACM-Reference-Format}
\bibliography{references} 

\end{document}